# Tracking ultrafast hot-electron diffusion in space and time by ultrafast thermo-modulation microscopy

A. Block[1,*], M. Liebel[1], R. Yu[1], M. Spector[2], Y. Sivan[3], F.J. García de Abajo[1,4], N.F. van Hulst[1,4,*]

[1] ICFO - Institut de Ciencies Fotoniques, The Barcelona Institute of Science and Technology, 08860 Castelldefels (Barcelona), Spain

[2] Department of Physics, Ben-Gurion University, 8410501 Be'er Sheva, Israel

[3] Unit of Electrooptics Engineering, Ben-Gurion University of the Negev, 8410501 Be'er Sheva, Israel

[4] ICREA - Institució Catalana de Recerca i Estudis Avançats, 08010 Barcelona, Spain

The ultrafast response of metals to light is governed by intriguing non-equilibrium dynamics involving the interplay of excited electrons and phonons. The coupling between them gives rise to nonlinear diffusion behavior on ultrashort timescales. Here, we use scanning ultrafast thermo-modulation microscopy to image the spatio-temporal hot-electron diffusion in a thin gold film. By tracking local transient reflectivity with 20 nm and 0.25 ps resolution, we reveal two distinct diffusion regimes, consisting of an initial rapid diffusion during the first few picoseconds after optical excitation, followed by about 100-fold slower diffusion at longer times. We simulate the thermo-optical response of the gold film with a comprehensive three-dimensional model, and identify the two regimes as hot-electron and phonon-limited thermal diffusion, respectively.

Understanding light-induced charge-carrier transport is of vital importance in modern technological applications, such as solar cells, artificial photosynthesis, optical detectors, and heat management in optoelectronic devices[1–4]. For example, the pathway of energy transfer in solar cells directly influences their energy conversion efficiency, as charge carriers need to diffuse towards extraction regions before being lost to other decay channels[5]. Indeed, much active research on emerging sensitized or thin-film photovoltaics aims at engineering effective carrier transport, while preventing hot-carrier loss[6–8]. In modern optoelectronic devices, hot electrons are first accelerated by the incident field before they scatter and dissipate their energy on the nanometer to micrometer scale, eventually transferring excess energy either to the lattice or being harvested by Schottky junctions[9–12]. Over the last decades, pump-probe techniques have been widely used to study the ultrafast response of carrier and heat transport in different materials under optical excitation[13,14]. While these time-resolved studies have uncovered numerous aspects of ultrafast carrier dynamics, the nanoscale spatial transport has remained largely unexplored. Indeed, direct real-space mapping to watch the carriers diffuse both in space and time is challenging. Recently, the combination of ultrafast spectroscopy and nano-imaging techniques, such as electron microscopy, X-ray diffraction, or near-field microscopy, has opened up new avenues to visualize microscopic transport[15–18]. Unfortunately, these methods are invasive, require vacuum, or involve slow probe scanning. All-optical far-field pump-probe microscopy is an interesting alternative as it probes the system of interest in a perturbation-free manner while achieving both high temporal and spatial resolution[19,20]. Using concepts of super-resolution and localization microscopy[21], the spatio-temporal dynamics can be resolved beyond the diffraction limit by precisely measuring the nanometer spatial changes of an initially excited, diffraction-limited region[22]. Due to the relatively facile implementation, ultrafast microscopy is now emerging as an important tool to study exciton diffusion in semiconductors, molecular solids, and 2D materials[23–29]. In this context, the ultrafast carrier diffusion dynamics in noble metals, such as gold, is of particular importance for heat management in nanoscale devices, as well as femtosecond laser ablation, but has thus far only been studied in the time domain[30-44].

In this work, we utilize novel ultrafast microscopy to gain insight into the complex non-equilibrium dynamics of hot electrons in gold. When a metal interacts with light, its electrons are excited above the Fermi level (Figure 1a). These initially excited electrons trigger a cascade of cooling mechanisms that involve energy transfer between different metal subsystems, of which conduction electrons and lattice vibrations are the dominant ones. Indeed, the ultrafast thermal response is commonly described by considering the conduction band electrons and the ionic lattice as separate thermodynamic subsystems with distinct thermal properties, such as heat capacity and thermal conductivity[45]. Since the heat capacity of the electrons is substantially smaller than that of the lattice, the electron subsystem quickly reaches a high temperature Fermi-Dirac distribution, to which we refer to as "hot electrons", while the lattice stays close to ambient temperature under the excitation intensities considered here. The electrons subsequently cool and thermalize with the lattice within a few picoseconds. The electron relaxation and thermalization of the two subsystems is a direct result of an interplay between electron-phonon coupling and hot-electron diffusion[39]. Purely time-resolved studies typically cannot separate these two contributions and have consistently neglected lateral heat flow, which becomes particularly important when considering nanoscale systems[31,33,34]. Directly resolving hot-electron diffusion, a crucial step for understanding the ultrafast heat dynamics, has thus far not been possible due to a lack of spatio-temporal resolution[37,39,40,42,46].

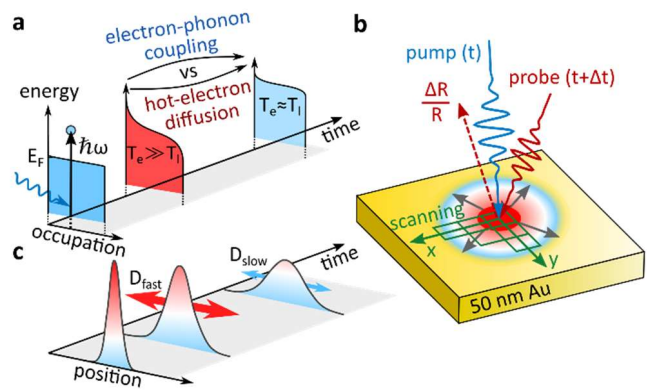

**Figure 1: Schematic description of scanning ultrafast thermo-modulation microscopy (SUTM).** (**a**) The energy distribution of the conduction band electrons at ambient temperature is perturbed by optical excitation. It quickly evolves to a quasi-thermalized "hot-electron" Fermi-Dirac distribution with high electron temperature ($T_e$), while the lattice temperature ($T_l$) stays close to the ambient level. Subsequent cooling due to electron-phonon coupling and hot-carrier diffusion leads to thermal equilibrium between the electron and lattice subsystems. (**b**) An optical pump pulse illuminates a 50 nm thin gold film, thus inducing a local hot-electron distribution. The probe pulse measures the temperature-dependent transient reflectivity ($\Delta R/R$) as a function of both pump-probe time delay and pump-probe spatial offset. (**c**) We monitor the spatio-temporal evolution of the photo-induced $\Delta R/R$ spot size with 20 nm and 250 fs resolution to visualize and distinguish hot-electron diffusion and thermal (phonon-limited) diffusion.

Here, we address this shortcoming by interrogating thin gold films with our recently developed scanning ultrafast thermo-modulation microscope (SUTM). We directly measure the spatio-temporal evolution of a locally induced hot-electron distribution on nanometer length scales with femtosecond resolution. Specifically, in our experiment, an optical pump pulse illuminates a thin gold film, thus creating hot carriers in the metal (Figure 1b). We measure the subsequent thermal response of the metal by employing a probe pulse that interrogates the sample at a well-defined time delay, $\Delta t$, with respect to the pump pulse. By spatially raster-scanning the probe beam relative to the stationary, tightly focused pump spot at each time delay, we obtain spatio-temporally-resolved transient reflection ($\Delta R/R$) maps. In this experiment, spatial heat diffusion manifests itself as a broadening of the initially excited area, which is quantified with a nanometer resolution far beyond the diffraction limit, by accurate determination of the SUTM spatial-response function (Figure 1c). Considering the electron redistribution dynamics described above (Figure 1a), we expect to identify distinct diffusion dynamics regimes, each dominated by a different diffusion mechanism, depending on the state of thermalization of the sample.

## Results

We image the light-induced thermal dynamics of a 50 nm thin gold film using SUTM. The sample is optically excited with a 450 nm (2.76 eV) pump pulse and interrogated with a 900 nm (1.38 eV) probe pulse, as outlined above (Figure 1b). We focus the beams to spots of full-width at half-maximum (FWHM) at the sample of 0.6 µm and 0.9 µm, respectively, and record $\Delta R/R$ maps at different times after photo-excitation by varying the pump-probe delay between -5 ps and 30 ps.

Figure 2a shows transient reflection images, recorded at three different pump-probe time-delays $\Delta t$ for a fixed pump fluence $F$ of 1.0 mJ/cm$^2$, which is well below the damage and ablation threshold of the metal[47]. At $\Delta t$ = -2 ps (i.e., when the probe interrogates the sample before photo-excitation by the pump), the $\Delta R/R$ response is negligible. Then, at $\Delta t$ = 0 ps, a negative $\Delta R/R$ spot emerges around the pump beam position $x = y = 0$, with up to -10$^{-4}$ contrast. Subsequently, at $\Delta t$ = 10 ps, we observe a reduced response, as the signal has decayed. The negative sign of $\Delta R/R$ indicates that the heated area exhibits decreased reflection. To investigate the temporal decay dynamics, we spatially overlap the pump and probe beams ($x = y = 0$) and vary the time-delay. Figure 2b shows the resulting trace. The transient reflection shows a negative step response with a 300 fs rise time, close to our instrument temporal resolution, followed by a biexponential decay with fast (1 ps) and slow (930 ps) components (see Supplementary Fig. 1). While this data contains information about the temporal carrier dynamics, the spatial diffusion information is provided by the $\Delta R/R$ maps of Figure 2a. Therefore, we fit the central cross sections of the images with Gaussian functions (Figure 2c). The FWHM increases by about 100 nm from 1.05 µm at $\Delta t$ = 0 ps to 1.15 µm at $\Delta t$ = 10 ps, a result that we attribute to spatial heat diffusion. The accuracy of this method is ultimately limited by the signal-to-noise ratio of the response function, which dictates how well the profiles can be fit. We observe a FWHM accuracy of about 20 nm.

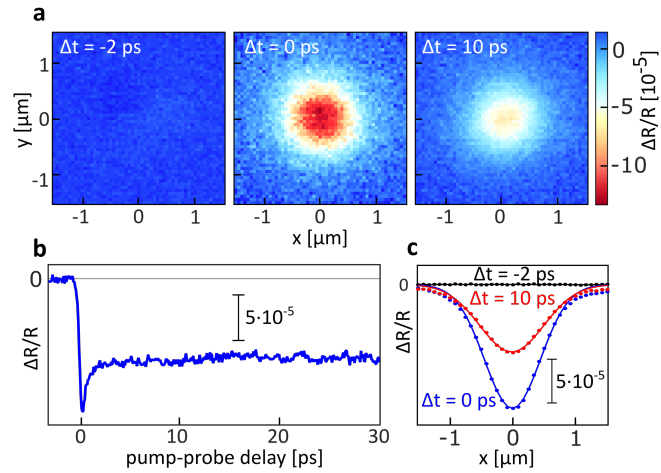

**Figure 2: Hot-electron dynamics.** (**a**) Scanning ultrafast thermo-modulation images of the gold film recorded by spatially scanning the probe beam ($\lambda_{probe}$ = 900 nm) relative to a fixed pump beam position ($\lambda_{pump}$ = 450 nm, centered at $x = y = 0$ in the image) for selected pump-probe delays. (**b**) Transient reflectivity dynamics for collinear pump and probe pulses, exhibiting two distinct exponential decay contributions with time constants of 1 ps and 930 ps, respectively. An offset at $\Delta t$ < 0 has been subtracted from the data. (**c**) Spatial profiles (dots) and Gaussian fits (curves) for three selected pump-probe delays, extracted from **a** by cutting horizontal lines through the center of the spots ($y = 0$).

We further investigate this evident diffusion behavior, as we want to track the rate at which it takes place during the thermalization process. Therefore, we proceed to record transient reflectivity line profiles over many pump-probe delays. Figure 3a shows a typical spatio-temporal dataset for $F$ = 1.0 mJ/cm$^2$. We obtain an estimate of the time-dependent diffusion coefficient $D(t)$ in a first, semi-quantitative manner by assuming the following relationship between the width (FWHM) and $D$ for an initial Gaussian profile, which we adopt from a general treatment of diffusion problems[48] (see Supplementary Information (SI) for details):

$$\frac{\partial \text{FWHM}^2(t)}{\partial t} = 16 (\ln 2) D(t). \quad (1)$$

We extract the FWHM at each time-delay by fitting Gaussian profiles to the $\Delta R/R$ maps, as previously explained, and plot the resulting temporal evolution of the squared width, FWHM$^2$, in Figure 3b. We observe an initial fast spreading, revealed by an increase in FWHM$^2$, followed by a much slower broadening at longer time delays. The two diffusion regimes appear to correlate with the fast and slow temporal decay regimes of $\Delta R/R$. We estimate the initial and final diffusion coefficients by fitting lines to the curve in Figure 3b (Equation 1). Excluding the transition region, we restrict our fit to the $\Delta t$ = 0 - 1 ps and 5 - 30 ps intervals, yielding $D_{\text{fast}}$ = 95 cm$^2$/s and $D_{\text{slow}}$ = 1.1 cm$^2$/s, respectively.

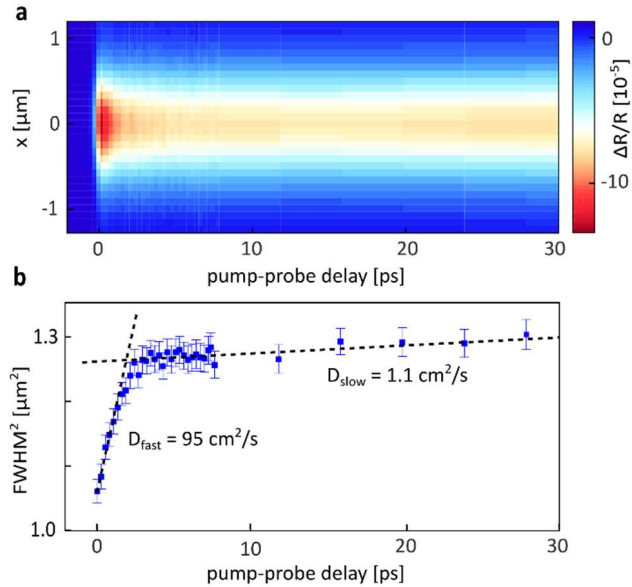

**Figure 3: Two-step diffusion dynamics.** (**a**) Spatio-temporal dynamics of the transient reflection signal $\Delta R/R$. We scan the probe beam over the pump beam (1D scan across spot center, vertical axis) as a function of pump-probe delay (horizontal axis). The offset at $\Delta t$ < 0 has been subtracted from the data. (**b**) Squared-width evolution of the $\Delta R/R$ profile, extracted by Gaussian fitting to the spatial profile at each pump-probe delay (symbols). The error bars show the 68% confidence intervals of the Gaussian fits. We extract the initial and final diffusion coefficients by fitting slopes in the two regions (dashed lines) and comparing to Equation 1. We find fast diffusion of $D_{\text{fast}}$ = 95 cm$^2$/s within the first few picoseconds, followed by slower diffusion ($D_{\text{slow}}$ = 1.1 cm$^2$/s) after >5 ps.

The above analysis provides a simple first measure to quantify diffusion. Yet, it relies on the assumption of a single diffusing profile and its proportionality to transient reflection, while ignoring the underlying electron and phonon subsystems, as well as their individual thermal contributions to the reflection signal. To fully understand the nature of the time-dependent diffusion mechanisms at work, in particular the transition between the two regimes of the observed spot broadening dynamics, it is necessary to model the response of the system more rigorously. Briefly, we model the spatio-temporal evolution of the pump-induced changes to the reflectivity of the sample in three basic steps (Figure 4a). First, we calculate the electron and lattice temperature distributions in the gold film, $T_e(\mathbf{r},\Delta t)$ and $T_l(\mathbf{r},\Delta t)$ ($\mathbf{r} = x,y,z$), resolved in space and time, by means of a 3D two-temperature model[43,45]. We then calculate the spatio-temporal dynamics of the gold film permittivity $\varepsilon(\mathbf{r},\Delta t)$, considering the effect of both temperatures on the Drude response. Finally, we convert the permittivity to our observable, transient reflection $\Delta R/R(x,y,\Delta t)$, using the Fresnel equations (see Methods and Supplementary Fig. 2). This procedure allows us to directly compare the measured data to the predictions of our model for the spatio-temporal diffusion by Gaussian fitting of the resulting $\Delta R/R$ maps to obtain FWHM$^2$ as a function of $\Delta t$.

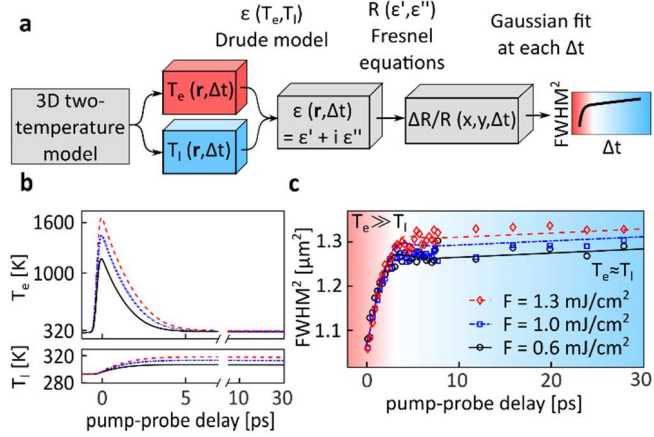

**Figure 4: Theoretical modeling and identification of different diffusion regimes.** (**a**) We simulate the spatio-temporal evolution of the optically excited gold film with a full 3D-space two-temperature model. We obtain the temperature-dependent complex permittivity from the calculated spatio-temporal electron and lattice temperature maps, including the thermal dependence of electron-electron and electron-phonon scattering, as well as thermal expansion of the lattice. We then calculate $\Delta R/R(x,y,\Delta t)$ using the thin-film Fresnel equations and extract its spatial dynamics using the same Gaussian fitting as in the experimental data analysis. (**b**) Predicted temporal evolution of the electron (top) and lattice (bottom) temperatures at the beam center for the three pump fluences used in the experiment. (**c**) Theoretical (curves) and experimental (symbols) evolution of the squared width of $\Delta R/R$ for different pump fluences $F$ (same as in **b**). In accordance with Figure 3b, we identify a fast diffusion regime at high electron temperatures, within the first few ps, followed by a thermalized regime (> 5 ps) dominated by phonon-limited transport, with orders-of-magnitude lower diffusivity.

In more detail, we describe our system with a two-temperature model (i.e., in terms of two thermodynamic subsystems, the electrons and the lattice[45]). We assume thermalization of the electron subsystem, as our temporal resolution of 250 fs cannot resolve the initial non-thermal stage right after photo-excitation. We model the 3D spatio-temporal temperatures in the gold film, $T_e(\mathbf{r},\Delta t)$ and $T_l(\mathbf{r},\Delta t)$, respectively, by considering the following coupled equations for the energy exchange between the gold electrons, the gold lattice, and the glass substrate, driven by a source $S(\mathbf{r},t)$ (the optical pump):

$$C_e(T_e)\frac{\partial T_e}{\partial t} = \nabla(k_e(T_e)\nabla T_e) - G(T_e - T_l) + S - S_{es}, \tag{2a}$$

$$C_l\frac{\partial T_l}{\partial t} = \nabla(k_l \nabla T_l) + G(T_e - T_l) - S_{ls}, \tag{2b}$$

$$C_s\frac{\partial T_s}{\partial t} = \nabla(k_s \nabla T_s) + S_{ls} + S_{es}, \tag{2c}$$

where $C_e/C_l$ ($C_s$) and $k_e/k_l$ ($k_s$) are the volumetric heat capacity and thermal conductivity of the gold electron/lattice (glass substrate), respectively; $G$ is the electron-phonon coupling coefficient; and $S_{es}$ ($S_{ls}$) is an energy exchange parameter at the gold-glass interface for the gold electrons (lattice), as described in Ref. 49 (see SI for more details, and Methods for the parameter values). We define the source term $S(\mathbf{r},t)$ for the energy absorbed from the pump pulse via its Gaussian spatial and temporal profiles (details in SI). Importantly, we are careful to include the lateral ($x,y$) temperature gradient, as we are interested in the prediction of the model for the lateral heat flow out of the focal volume. A finite element method calculation yields the 3D spatio-temporal electron and lattice temperatures $T_e(\mathbf{r},\Delta t)$ and $T_l(\mathbf{r},\Delta t)$ for the sample geometry. The calculated evolution of the electron and lattice temperatures $T_{e/l}$ at $x = y = z = 0$ is shown in Figure 4b. We observe a rapid increase in electron temperature to more than 1000 K, and a subsequent cooling and thermalization with the lattice temperature at few tens of degrees above the ambient level of 293 K. Next, we calculate the spatio-temporal dependence of the gold film permittivity $\varepsilon(T_e(\mathbf{r},\Delta t),T_l(\mathbf{r},\Delta t))$, using a Drude model for the near-infrared probe wavelength (see Methods). The model includes electron-electron Umklapp scattering, electron-phonon scattering, and thermal lattice expansion in the calculation of the gold permittivity, which therefore depends on both $T_e$ and $T_l$ (see SI). Finally, the pump-induced complex permittivity is converted into transient reflectivity using the Fresnel coefficients for thin films as a function of lateral position ($x,y$) and time ($\Delta t$) to yield $\Delta R/R(\varepsilon(\mathbf{r},\Delta t))$, assuming a uniform permittivity across the film thickness. The calculation actually predicts a higher $\Delta R/R$ contrast compared to experiment, possibly due to an overestimate of the absorbed power, however this does not affect the width distributions. We account for the effect of the finite spatial extension of the probe pulse by convolving the resulting $\Delta R/R$ map with its diffraction-limited width. Finally, we analyze the predicted spatio-temporal evolution of $\Delta R/R$ by Gaussian-fitting to the spatial cross sections, as described above for the experimental data. Figure 4c shows the calculated evolution of the squared width together with the experimental results for the three pump fluences under consideration. The comparison shows that the calculated evolution describes the experimental data well for both thermalization regimes, as well as the fluence dependence.

## Discussion

Using scanning ultrafast thermo-modulation microscopy, we have imaged the thermo-optical dynamics of gold films and identified two regimes of thermal diffusion dominated by hot-electrons and lattice modes (phonons), respectively. Intuitively, the two observed regimes of heat diffusion can be readily understood from limiting cases of the two-temperature model. At early times, when $T_e \gg T_l$, the electron-lattice thermalization time can be estimated as $\tau_{\text{e-ph}} = C_e(T_e)/G$, and the hot-electron-dominated diffusion is given by $D_{\text{fast}} = k_e/C_e$[30]. In the long term, after thermalization of the electrons and the lattice ($T_e \approx T_l$), the two-temperature model simplifies to a diffusion equation with $D_{\text{slow}} = (k_e+k_l)/(C_e+C_l) \approx k_e/C_l$. This is the well-known thermal diffusivity of gold[50]. Inserting the values of electron-phonon coupling constant, heat capacity, and thermal conductivity leads to $\tau_{\text{e-ph}} = 1 - 3$ ps (for $T_e = 300 - 1000$ K), $D_{\text{fast}} = 152$ cm$^2$/s and $D_{\text{slow}} = 1.36$ cm$^2$/s. Both these diffusion coefficients overestimate our observed values. Clearly, a solely temperature–based model is too simplistic. For a quantitative agreement with the experimental data, a full dynamics calculation of the spatial width of $\Delta R/R$ is needed, especially to include the transition between the two mentioned regimes. Our simulations of the full temperature development illustrate how the elevated temperature of the conduction-band electrons correlate with a fast carrier diffusion. Modeling the full dynamics from absorption through electron-lattice thermalization to permittivity spatio-temporal dynamics allows us to predict the evolution of the spatial width of $\Delta R/R$, resulting in remarkable quantitative agreement with the experimentally determined time-dependent diffusion. We note that our model relies purely on reported material constants and has no free adjustable parameters (see Methods). In addition, the dependence of the calculated width evolution on the laser fluence agrees well with the experimental data.

In the transition regime, around 5 ps time-delay, we predict a subtle decrease of FWHM, which lies within the uncertainty of the experimental data. In fact, the two-temperature-model predicts a substantial decrease of FWHM for the electron temperature distribution (i.e., apparent negative diffusion) in this transition regime. The effect is less visible after conversion to $\Delta R/R$, both experimentally and in simulation, presumably due to the influence of the lattice temperature in the crossover regime.

Recent experiments have studied the effects of non-thermal electron distributions, as well as a transition from ballistic to diffusive electron transport[31,33]. It will be interesting to study these effects in real space with SUTM at higher temporal resolution.

In summary, we have tracked thermally induced diffusion in a thin gold film from absorption to thermalization in time and space with femtosecond and nanometer resolution. We resolve, for the first time, a transition from hot-electron to phonon-limited diffusion. We interpret the electronic and phononic cooling regimes with the help of full two-temperature and thermo-optical modeling. The predicted dynamics agree well with the experimental observations. The insight gained on hot-carrier dynamics from direct spatio-temporal imaging is crucial to understand the interplay of electrons and phonons in ultrafast nanoscale photonics, and thus to design nanoscale thermal management in nano-optoelectronic devices, such as phase-change memory devices and heat assisted magnetic recording heads. Especially, the control of heat exchange between the electron and lattice systems is important in device functionality. More generally, the excess energy of hot electrons finds applications in an increasingly wide range of systems, such as thermoelectric devices, broadband photodetectors, efficient solar cells and even plasmon-enhanced photochemistry. Here we have applied our method to gold, the "gold" standard for many such applications of electron heat. Surely, ultrafast thermo-modulation microscopy is equally suitable to study a vast range of other materials and systems in the future.

## Methods

**Experimental details.** The sample is fabricated by thermal evaporation of 50 nm of gold onto a cleaned glass cover slip. The laser source is a Ti:Sapphire oscillator (Coherent Mira 900) emitting at 900 nm wavelength with a repetition rate of 76 MHz and 150 fs pulse duration. We frequency-double the laser using a BBO crystal to create the 450 nm pump beam. We compress the pump and probe pulses individually with two prism compressors made of fused silica and SF10 glass, respectively, and characterize the temporal resolution of the entire imaging system as the 10-90% cross-correlation rise time at the sample plane of 250 fs. Both beams are focused onto the sample with a 40×/NA 0.6 objective lens (Olympus LUC Plan FLN). We measure the beam profile at the sample plane by scanning a line edge of the gold film through the beam. We minimize and overlap the pump and the probe beam foci in the sample plane by iterative (de-)collimation and beam profiling, resulting in $FWHM_{pump}$ = 0.6 µm and $FWHM_{probe}$ = 0.9 µm values at the same $z$ position. We scan the probe beam over the pump with a two-axis mirror galvanometer system (Thorlabs GVS012). We modulate the pump beam with a chopper (Newport New Focus 3501) at 6.4 kHz and record transient reflection by long-wave pass filtering (Omega Optical 585ALP and 740AELP) of only the probe beam onto a balanced photodiode (Thorlabs PDB450C) and lock-in amplification (Stanford Research Systems SR830). A sketch of the setup is shown in Supplementary Fig. 3.

**Two-temperature model (TTM), finite element method implementation.** For the TTM calculations (Eqs. (2)), we used the following parameters and scaling laws: the electronic volumetric heat capacity of gold[39] $C_e(T_e) = \gamma T_e$, with $\gamma$ = 71 Jm$^{-3}$K$^{-2}$; the electronic thermal conductivity of gold[39] $k_e(T_e,T_l) = k_0 T_e/T_l$, with $k_0$ = 317 Wm$^{-1}$K$^{-1}$; the lattice heat capacity and thermal conductivity of gold[51,52] $C_l$ = 2.327×10$^6$ Jm$^{-3}$K$^{-1}$ and $k_l$ = 2.6 Wm$^{-1}$K$^{-1}$, respectively; the electron-phonon coupling constant[53] $G$ = 2.2×10¹⁶ Wm⁻³K⁻¹; and the heat capacity and thermal conductivity of glass[54] $C_s$ = 1.848×10⁶ Jm⁻³K⁻¹ and $k_s$ = 0.8 Wm⁻¹K⁻¹, respectively. See SI for further details.

**Thermo-optical response calculation.** We model the complex permittivity at the infrared probe wavelength with a Drude model permittivity[55]

$$\varepsilon = \varepsilon_\infty - \frac{\omega_p^2(T_l)}{\omega(\omega + i\gamma_{re}(T_e, T_l))}.$$

Here, $\varepsilon_\infty$ = 9.5 is the high-frequency permittivity[55]. The plasma frequency $\omega_p$ depends on lattice temperature $T_l$ due to volume expansion affecting the free conduction band electron density $n_e$. Namely, $\omega_p(T_l) = \sqrt{e^2/\varepsilon_0 m_{\text{eff}} \times n_e(T_0)/(1 + \beta \Delta T_l)}$, where $\beta = 4.23 \times 10^{-5}$ K⁻¹ is the thermal expansion coefficient[56], $n_e(T_0)$ = 5.9×10²² cm⁻³ is the unperturbed density[57], $m_{\text{eff}} \approx m_e$ is the effective electron mass, $\varepsilon_0$ is the vacuum permittivity, and $m_e$ and $e$ and are the elementary charge and mass, respectively. Further, $\gamma_{re}$ represents the total rate of relaxation collisions that conserve momentum and energy of the electron subsystem, given by $\gamma_{re}(T_e, T_l) = \gamma_{e-ph}(T_l) + \gamma_{e-e}^{Um}(T_e)$, where $\gamma_{e-ph}$ is the electron-phonon collision rate, depending on the lattice temperature as $\gamma_{e-ph}(T_l) = B T_l$, and $\gamma_{e-e}^{Um}$ is the Umklapp electron-electron collision rate, depending on $T_e$ as $\gamma_{e-e}^{Um}(T_e) = \Delta^{(Um)} A T_e^2$, with A = 1.7×10⁷ K⁻²s⁻¹, B = 1.45×10¹¹ K⁻¹s⁻¹ and $\Delta^{Um}$ = 0.77[58,59] (see SI for details).

## Data Availability

The data that support the findings of this study are available from the corresponding author on request.

## Acknowledgements

A.B. acknowledges financial support from the International PhD fellowship program 'la Caixa'—Severo Ochoa. We thank P. Woźniak for helpful discussions while preparing the manuscript. This project is financially supported by the European Commission (ERC Advanced Grants 670949-LightNet and 789104-eNANO), Spanish Ministry of Economy ("Severo Ochoa" program for Centres of Excellence in R&D SEV-2015-0522, PlanNacional FIS2015-69258-P and MAT2017-88492-R, and Network FIS2016-81740-REDC), the Catalan AGAUR (2017SGR1369), Fundació Privada Cellex, Fundació Privada Mir-Puig, and Generalitat de Catalunya through the CERCA program.

## Author information

**Affiliations**

*ICFO - Institut de Ciencies Fotoniques, The Barcelona Institute of Science and Technology, 08860 Castelldefels (Barcelona), Spain*

Alexander Block, Matz Liebel, Renwen Yu, F. Javier García de Abajo & Niek F. van Hulst

*Department of Physics, Ben-Gurion University, Be'er Sheva, Israel*

Marat Spector

*Unit of electro-optics Engineering, Ben-Gurion University, Be'er Sheva, Israel*

Yonatan Sivan

*ICREA - Institució Catalana de Recerca i Estudis Avançats, 08010 Barcelona, Spain*

Javier García de Abajo & Niek F. van Hulst

**Contributions**

A.B., with the help of M.L., built the experiment, fabricated the samples, performed the measurements, and analyzed the data. A.B., R.Y., M.S., Y.S., and F.J.G.d.A. contributed to the theoretical modeling. M.L. and N.F.v.H. conceived the experiment. A.B. wrote the manuscript. All authors contributed to the interpretation the data, discussion and writing of the manuscript.

**Competing Interests**

The authors declare no competing interests.

**\*Corresponding authors**

Correspondence to Alexander Block or Niek F. van Hulst

Alexander.Block@ICFO.eu, Niek.vanHulst@ICFO.eu

Supplementary Information for:

# Tracking ultrafast hot-electron diffusion in space and time by ultrafast thermomodulation microscopy

A. Block, M. Liebel, R. Yu, M. Spector, Y. Sivan, F.J. García de Abajo, N.F. van Hulst

**Supplementary note 1: Two-temperature model**

In this model, the electron and lattice temperature $T_e$ and $T_l$ are assumed to be time- and position-dependent, with their spatio-temporal behavior being described by the following coupled equations [1]:

$$C_e(T_e)\frac{\partial T_e}{\partial t} = \nabla(k_e(T_e)\,\nabla T_e) - G(T_e - T_l) + S - S_{es} \qquad (1a)$$

$$C_l \frac{\partial T_l}{\partial t} = \nabla(k_l\,\nabla T_l) + G(T_e - T_l) - S_{ls} \qquad (1b)$$

$$C_s \frac{\partial T_s}{\partial t} = \nabla(k_s\,\nabla T_s) + S_{ls} + S_{es} \qquad (1c)$$

with $C_e(T_e) = \gamma\,T_e$, $\gamma = 71$ Jm$^{-3}$K$^{-2}$, for the electron heat capacity [2] ; $C_l = 2.45 \cdot 10^6$ Jm$^{-3}$ K$^{-1}$ for the lattice heat capacity [3]; $k_e(T_e,T_l) = k_0\,T_e/T_l$, with $k_0 = 317$ Wm$^{-1}$K$^{-1}$, for the electron thermal conductivity [2]; $k_l = 2.6$ Wm$^{-1}$K$^{-1}$, for the lattice thermal conductivity [4]; $C_s = 1.848 \cdot 10^6$ Jm$^{-3}$ K$^{-1}$ for the glass substrate heat capacity [5] ; $k_s = 0.8$ Wm$^{-1}$K$^{-1}$ for the substrate thermal conductivity [5] ; $G = 2.2 \cdot 10^{16}$ Wm$^{-3}$K$^{-1}$ for gold electron-phonon coupling coefficient [6]; $S_{es/ls} = G_{es/ls}(T_{e/l} - T_s)$ represents the boundary interface heat exchange between the gold electron/lattice and the glass substrate, where $G_{es} = (96.12 + 0.189\,T_e)$ MWm$^{-2}$K$^{-1}$ and $G_{ls} = 141.5$ MWm$^{-2}$K$^{-1}$ are the corresponding thermal boundary conductance, respectively [7]. In Eq. (1a), the source term is defined for a Gaussian beam excitation as [8]

$$S(\mathbf{r},t) = \frac{S_a}{\delta} F \cdot \exp\left[-\frac{z}{\delta} - (4\ln 2)\frac{x^2 + y^2}{\text{FWHM}_{\text{pump}}^2}\right] \cdot \frac{1}{t_{\text{pump}}} \sqrt{\frac{4\ln 2}{\pi}} \exp\left[-(4\ln 2)\frac{(t - 2t_{\text{pump}})^2}{t_{\text{pump}}^2}\right], \qquad (2)$$

with a full-width at half-maximum $\text{FWHM}_{\text{pump}} = 0.6$ µm, and a pulse duration $t_{\text{pump}} = 0.2$ ps. The absorbance $S_a = 1 - R - T = 0.5$ and the penetration depth $\delta = 18.7$ nm are estimates based on tabulated optical data for gold at the pump wavelength of 450 nm [9], with $R$ and $T$ being the thin film reflectivity and transmissivity, respectively. $F$ is the pump laser fluence.

**Supplementary note 2: Thermo-optical response**

Here, we describe the model used to describe the electron and lattice temperature dependence of the permittivity (or dielectric function) ε.

The dielectric function of the gold film as a function of electron and lattice temperature at the probe wavelength ($\lambda_{\text{probe}} = 900$ nm) is well described by the Drude model [9], in which electrons are considered as free charges moving in response to an optical field oscillating at angular frequency ω. Using a classical equation of motion, the dielectric response of the plasma was shown to be

$$\varepsilon_{\text{intra}} = \varepsilon_\infty - \frac{\omega_p^2(T_l)}{\omega(\omega + i\gamma_{re}(T_e,T_l))} \qquad (3)$$

with the plasma frequency $\omega_p(T_l)$ being

$$\omega_p(T_l) = \sqrt{\frac{e^2}{\epsilon_0 m_{eff}} \cdot n_e(T_l)}. \qquad (4)$$

Here, $e$ is the electron charge, $n_e$ is the free electron density in the conduction band ($n_e = 5.9 \cdot 10^{22}$ cm$^{-3}$ for Au [10]), $\varepsilon_0$ is the dielectric permittivity of the vacuum, $m_{\text{eff}}$ is the effective mass (estimated as equal to the electron mass, $m_e$), and $\gamma_{re}$ represents the total rate of collisions that conserve the total momentum and/or energy of the electron subsystem, i.e., collisions that cause relaxation of the electron system (in contrast to like regular e-e collisions, which affect the permittivity only indirectly, by causing a thermalization, i.e., a temperature increase). It is given by:

$$\gamma_{re}(T_e, T_l) = \gamma_{e-ph}(T_l) + \gamma_{e-e}^{Um}(T_e), \qquad (5)$$

where the subscript *re* stands for relaxation, $\gamma_{e-ph}$ is the e-ph collision rate and $\gamma_{e-e}^{Um}$ is the Umklapp e-e collision rate.

$\varepsilon_\infty$ represents the off-resonant (i.e., high frequency) contribution of the interband transitions ($\varepsilon_\infty$ = 9.5 for Au [9]). Here, we neglect its dependence on the temperatures which, according to recent results [11,12], is weak at the 900nm probe wavelength. In addition, note that for 1200 K $\gg$ $T_e$ > $T_{Debye}$, $T_l \approx$ 300 K, one can neglect the temperature dependence of $m_{eff}$ [13]. Thus, heating of a metal affects the intraband contribution to the dielectric constant only via two parameters, the plasma frequency $\omega_p$ and the constituents of the electron relaxation rate $\gamma_{re}$. We now specify in detail the temperature dependence of these parameters. The e-ph scattering rate, $\gamma_{e-ph}$, is given by $\gamma_{e-ph}(T_l) = B\, T_l$, where for Au, B ≈ 1.45 · 10$^{11}$ K$^{-1}$s$^{-1}$ [14].

$\gamma_{e-e}^{Um}$ represents the intraband transition rate due to Umklapp e-e scattering events[1]. As shown by Kaveh and Wiser [15], within the framework of Fermi Liquid Theory [16], it equals the fraction $\Delta^{(Um)}$ of the total e-e scattering rate, $\gamma_{e-e}$, i.e.,

$$\gamma_{e-e}^{Um}(T_e) = \Delta^{(Um)}\gamma_{e-e}, \tag{6}$$

where $\gamma_{e-e} = AT_e^2$, and $A = K(\pi k_B)^2/2$. $K = m_e^3 W_{e-e}/(8\pi\hbar^6)$ is the characteristic e-e scattering constant that contains the angular-averaged scattering probability $W_{e-e}$ [16]; $k_B$ is the Boltzmann constant. For Au, A ≈ 1.7 · 10$^7$ K$^{-2}$s$^{-1}$ [14] and $\Delta^{(Um)}$= 0.77 [17]. Thus, $\gamma_{e-e}^{Um}$ depends quadratically on $T_e$ [18].

The plasma frequency is affected by the temperature rise via a change of $n_e$ due to volume expansion of the metallic system[2] [19]

$$V(T_l) = V_{eq}(1 + \beta \Delta T_l), \tag{7}$$

where $V_{eq}$ is the equilibrium volume of the metallic system at ambient temperature $T_0$ and $\beta$ is the volume thermal expansion coefficient[3]. We assume that the total number of electrons in the conduction band, $N_e$, is independent of temperature

$$N_e = n_e(T_0)V_{eq} = n_e(T_l)V(T_l), \tag{8}$$

so that the density is given by

$$n_e(T_l) = \frac{n_e(T_0)}{1+\beta\Delta T_l}. \tag{9}$$

Substitution of Eq. (9) in Eq. (4) leads to [19,20]

$$\omega_p(T_l) = \frac{\omega_p(T_0)}{\sqrt{1+\beta\Delta T_l}}, \tag{10}$$

---

[1] In *normal e-e* scattering, the momentum is conserved so that it does not contribute directly to the intraband collision rate, $\gamma_{re}$. However, these collisions do cause the electron system to thermalize, hence, the electron temperature to increase, so that they contribute to the permittivity indirectly through the temperature dependence of other quantities in Eq. (2). *Umklapp e-e* scattering processes, however, impart momentum to the lattice as a whole, thus, the total electron momentum is not conserved, whereas the quantity that is conserved is the quasi-momentum. Violation of momentum conservation leads to contribution to the collision rate [23].

[2] For the near-infrared probe wavelength, one can safely neglect the possibility of the density of the electrons in the conduction band to increase due to interband transitions.

[3] The thermal expansion coefficient is directly related to the Gruneisen constant $\gamma_G$ according to the following equation [24]: $\beta \approx \gamma_G C_l/B_b$, where $B_b$ is the bulk modulus. Using the data for $\gamma_G$ and for bulk modulus $B_b$ for Au [25], the thermal expansion coefficient is set to β = 4.23·10$^{-5}$ K$^{-1}$.

or

$$\omega_p(T_l) = \sqrt{\frac{e^2}{\epsilon_0 m_{eff}} \cdot \frac{n_e(T_0)}{1+\beta \Delta T_l}}. \tag{11}$$

**Supplementary note 3: Calculation of the transient reflectivity**

From the spatiotemporal maps of $T_e(\mathbf{r},\Delta t)$ and $T_l(\mathbf{r},\Delta t)$, we first calculate the reflectivity $R(T_e,T_l)$, and then the transient reflectivity $\Delta R/R = [R(T_e,T_l) - R(T_0,T_0)]/R(T_0,T_0)$. We use the thin-film Fresnel equations [21] for the reflection by a thin film (thickness $h$, index of refraction $n_2 = \sqrt{\varepsilon}$ for gold [9]) under perpendicular incidence from air ($n_1 = 1$), with the film supported on a glass substrate ($n_3 = 1.5$).

$$R = |r|^2, \text{ with } r = \frac{r_{12} + r_{23} \cdot \exp(2i\beta)}{1 + r_{12} \cdot r_{23} \cdot \exp(2i\beta)}, \tag{12}$$

where $\beta = 2\pi n_2 h/\lambda_0$, $r_{12} = (n_1 - n_2)/(n_1 + n_2)$, $r_{23} = (n_2 - n_3)/(n_2 + n_3)$.

We note that as an alternative approach, we have also calculated $R$ from finite z-slices of the 3D permittivity $\varepsilon(x,y,z)$, extracted directly from the 3D finite element method calculation mesh $T_e(x,y,z)$ and $T_l(x,y,z)$, using the transfer matrix method. As this calculation yielded nearly the same results as the simpler version (Eq. 12, assuming homogeneous permittivity along the film thickness z), we trust the validity of this approach.

**Supplementary note 4: General diffusion model**

As stated in the main text, we estimate diffusivity, or diffusion coefficient, in a first approximation from the slope of the FWHM² vs. time plot, before the more rigorous modeling described above. Thus, to quantify a general diffusion process with a time-dependent diffusivity $D(t)$, we assume that the absorbed pump power induces a Gaussian energy distribution at the sample $u(\mathbf{x},t=0) \propto \exp(-4\ln 2\, |\mathbf{x}|^2/\text{FWHM}_{\text{pump}}^2)$, where $\mathbf{x} \in \mathbb{R}^n$ ($n$ = 1, 2, or 3, spatial dimension) and that $u$ evolves according to a heat equation [22],

$$\frac{\partial u(\mathbf{x},t)}{\partial t} = D(t)\nabla^2 u(\mathbf{x},t), \tag{13}$$

or, alternatively, a heat equation including a decay with a rate $\Gamma = 1/\tau$

$$\frac{\partial u(\mathbf{x},t)}{\partial t} = D(t)\nabla^2 u(\mathbf{x},t) - \Gamma\, u(\mathbf{x},t). \tag{14}$$

Interestingly, both solutions of $u(\mathbf{x},t)$ to equations (13) and (14) are temporally decaying spatial Gaussian profiles with a time-dependent width. The full-width at half-maximum, measured in any one dimension, FWHM($t$), is related to $D(t)$ via[4]:

$$\frac{\partial \text{FWHM}^2(t)}{\partial t} = 16 \ln 2\, D(t). \tag{15}$$

Under the assumption that our observed $\Delta R/R(\mathbf{x},\Delta t)$ profiles scale linearly with $u(\mathbf{x},t)$, we can determine $D(t)$ from our space-time-resolved data. We note that this derivation is often shown for the n-dimensional mean square displacement <$\mathbf{x}^2$>, which is dependent on the dimensionality $n$, yet the projection onto one axis, or a 1D "cutline" of the profile, is independent of n.

---

[4] This result is the generalization of the more common result known for a time-independent diffusivity $D$: $\sigma^2(t) = \sigma^2(0) + 2D \cdot t$, and converting the Gaussian width parameter σ to FWHM using $\text{FWHM} = 2\sqrt{2\ln 2}\,\sigma$.

**Supplementary Figure 1:**

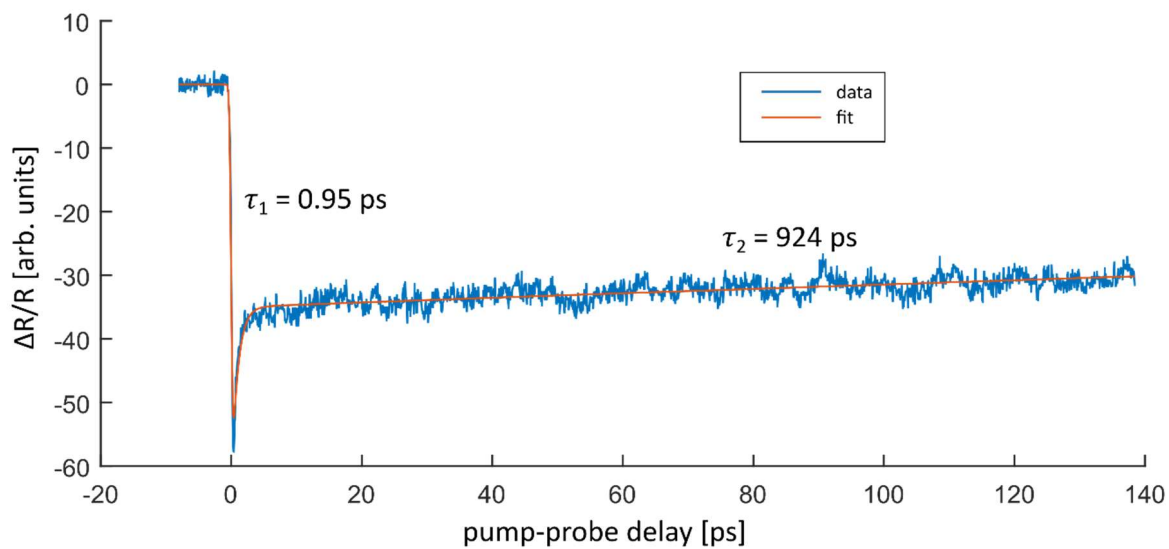

**Suppl. Fig. 1a:** Long-time dynamics of Figure 2b from the main text. The red line shows the biexponential fit to extract the two decay parameters.

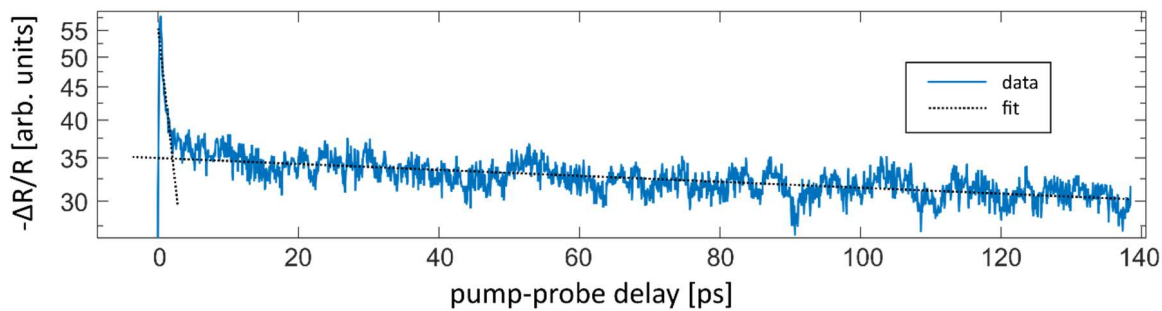

**Suppl. Fig. 1b:** Same as Suppl. Fig. 1a, on linear-log scale. Dashed lines show the exponential decay parameters taken from the above fit.

**Supplementary Figure 2:**

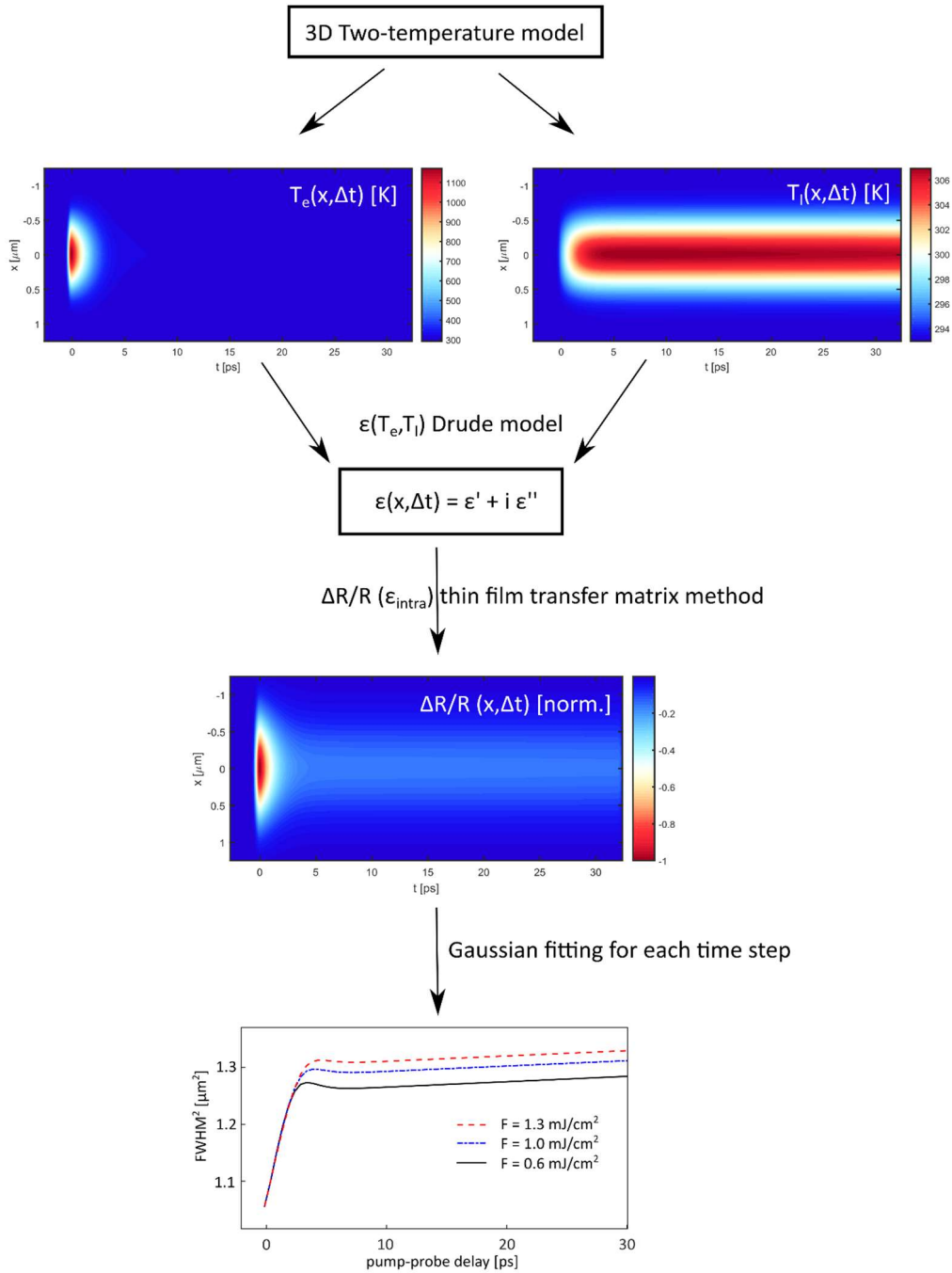

**Suppl. Fig. 2: Scheme of our calculations.** We calculate the temporal evolution of the spatial temperature profiles for the electron and lattice subsystems after optical excitation. Next, we calculate the spatio-temporal permittivity from the two-temperature maps using a temperature dependent Drude model described in the thermo-optical response section. Finally, we convert the resulting permittivity to a transient reflectivity using the Fresnel coefficients for thin films. The evolution of the spatial width is calculated by fitting the $\Delta R/R$ maps to Gaussians.

**Supplementary Figure 3:**

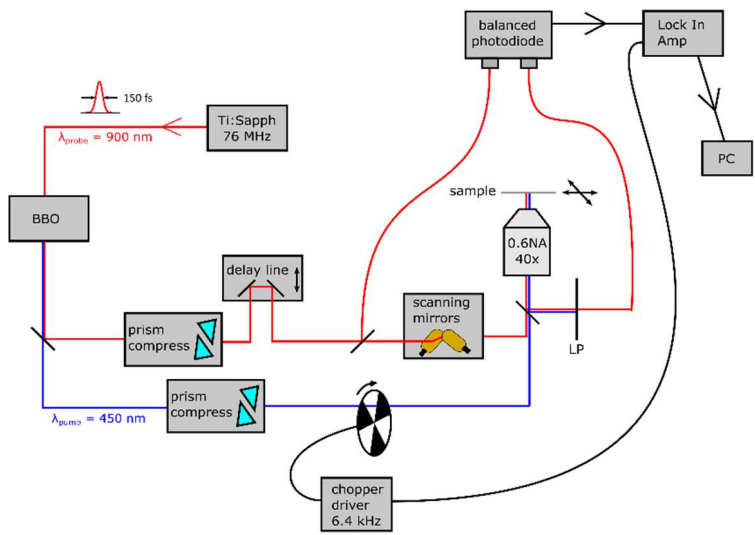

**Suppl. Fig. 3:** Schematic of the scanning ultrafast thermomodulation microscope. BBO: β-Barium borate crystal. The description of the experiment is found in the main text.

**References for Supplementary Information**